\documentclass[12pt]{article}
\usepackage{graphicx}
\usepackage[utf8]{inputenc}
\usepackage[T1]{fontenc}
\usepackage{indentfirst}
\usepackage[margin=0.6in,nomarginpar]{geometry}
\usepackage[final]{hyperref}
\usepackage{amsmath}
\usepackage{hyperref}
\usepackage{cite}

\hypersetup{
colorlinks=true,
linkcolor=blue,
citecolor=blue,
filecolor=magenta,
urlcolor=blue
}
\begin{document}

\title{Ideal Bose Gas and Blackbody Radiation in the Dunkl Formalism}
\author{F. Merabtine\thanks{%
merabtinefateh172@gmail.com} \\
Laboratory for Theoretical Physics and Material Physics Faculty of Exact \\
Sciences and Informatics, Hassiba Benbouali University of Chlef, Algeria \ \
\ \ \ \, \and B. Hamil\thanks{%
hamilbilel@gmail.com} \\
D\'{e}partement de physique, Facult\'{e} des Sciences Exactes, \\
Universit\'{e} Constantine 1,Constantine, Algeria. \and B. C. L\"{u}tf\"{u}o%
\u{g}lu\thanks{%
bekir.lutfuoglu@uhk.cz (Corresponding author)} \\
Department of Physics, University of Hradec Kr\'{a}lov\'{e},\\
Rokitansk\'{e}ho 62, 500 03 Hradec Kr\'{a}lov\'{e}, Czechia. \and A. Hocine%
\thanks{%
a.hocine@univ-chlef.dz} \\
Laboratory for Theoretical Physics and Material Physics Faculty of Exact \\
Sciences and Informatics, Hassiba Benbouali University of Chlef, Algeria \ \
\  \and M. Benarous\thanks{%
m.benarous@univ-chlef.dz} \\
Laboratory for Theoretical Physics and Material Physics Faculty of Exact \\
Sciences and Informatics, Hassiba Benbouali University of Chlef, Algeria \ \
\ }
\date{\today }
\maketitle

\begin{abstract}
Recently, deformed quantum systems gather lots of attention in the
literature. Dunkl formalism differs from others by containing the
difference-differential and reflection operator. It is one of the most
interesting deformations since it let us discuss the solutions according to
the even and odd solutions. In this work, we studied the ideal Bose gas and
the blackbody radiation via the Dunkl formalism. To this end, we made a
liaison between the coordinate and momentum operators with the creation and
annihilation operators which allowed us to obtain the expressions of the
partition function, the condensation temperature, and the ground state
population of the Bose gas. We found that Dunkl-condensation temperature
increases with increasing $\theta$ value. In the blackbody radiation
phenomena, we found how the Dunkl formalism modifies total radiated energy.
Then, we examined the thermal quantities of the system. We found that the
Dunkl deformation causes an increase in entropy and specific heat functions
as well as in the total radiation energy. However, we observed a decrease in
the Dunk-corrected Helmholtz free energy in this scenario. Finally, we found
that the equation of state is invariant even in the considered formalism.
\end{abstract}


\section{Introduction}

It would not be wrong to say that the foundations of Dunkl formalism are
based on an article, which was published in the middle of the last century 
\cite{Wigner}. In this article, Wigner handled two systems, namely the free
case and the classical harmonic oscillator, and he tried to derive
commutation relations from the equation of motion. Surprisingly, his
calculations ended with an extra free constant which meant that the
commutation relations cannot be obtained uniquely in the examined systems.
In 1951, Yang reworked this issue by considering the quantum harmonic
oscillator problem instead of the classical one under more precise
definitions of Hilbert space and series expansions \cite{Yang}. He found
that adding a reflection operator to the momentum operator would remove the
arbitrariness of the commutation relations. A few decades later, via a
purely mathematical point of view, Dunkl contributed to the ongoing
discussion on the relations between differential-difference and reflection
operators by describing a new derivative operator that could be used instead
of the usual partial derivative \cite{Dunkl1}. It is noteworthy that if the
Dunkl momentum operator is defined using the Dunkl derivative, an operator
similar to Yang's momentum operator is obtained. Therefore, Dunkl's
derivative achieved great attention not only in mathematics \cite{Rosler},
but also in physics \cite{Lapointe, Kakei, Klishevich, Horvathy, G1, G2, G3,
G4, G5, Ramirez1, Ramirez2, Sargol, Chung1, Ghaz, Mota1, Marcelo, Chungrev,
Kim, Jang, Ojeda, Mota2, Mota3, Merad, ChungDunkl, ChungEPJP, Hassan, Dong,
Bilel1, Bilel2, Seda, Mota2022, new1, new2, new3, new4}. Initially, the
interest of physicists was mainly based on the applicability of the
formalism in the Calogero-Sutherland-Moser models \cite{Lapointe, Kakei}. In
the last decade, relativistic and non-relativistic systems are being
examined within Dunkl-formalism. The main motivation in these studies is the
simultaneous derivation of the wave function solutions as odd and even
functions, thanks to the reflection operator. For example, in 2013, Genest
et al. studied the isotropic and anisotropic Dunkl-oscillator systems in the
plane, respectively in \cite{G1} and \cite{G2}. The next year, the same
authors discussed the Dunkl-isotropic system with algebraic methods in \cite%
{G3}, and they investigated solutions to the Dunkl-anisotropic system in
three dimensions \cite{G4}. In 2018, Sargolzaeipor et al. studied the Dirac
oscillator within the Dunkl formalism \cite{Sargol}. The next year, Mota et
al. solved the Dunkl-Dirac oscillator problem in two dimensions \cite{Mota1}. 
Recently, the Dunkl-Klein Gordon oscillator is examined in two and three
dimensions in \cite{Mota2, Mota3, Bilel1}. The Duffin-Kemmer-Petiau oscillator is also studied in the Dunkl formalism in \cite{Merad}.

The use of Dunkl's formalism in physics is not limited to these fields. For
instance, two of the authors of this manuscript investigated the thermal
properties of a graphene layer that is located in an external magnetic field
within the Dunkl formalism in\cite{Bilel2}. The nonrelativistic solution of
a position-dependent mass model is examined with a Lie algebraic method in 
\cite{Seda}. Recently, we observe papers that consider different
generalization forms of Dunkl derivative in the literature \cite{Mota2022,
new1, new2}. Dunkl formalism is also employed in noncommutative phase space
in \cite{new3}.

In a quite interesting paper, Ubriaco studied the thermodynamics of bosonic
systems associated with the Dunkl derivative from the statistical mechanical
point of view \cite{Marcelo}. He formulated a bosonic model on a Hamiltonian
that is defined in terms of Dunkl-creation and annihilation operators with
the help of $su(1,1)$ algebra and investigated the effect of the Wigner
parameter on the thermodynamic properties. He observed that the
Dunkl-critical temperature, the Dunkl-entropy, and the Dunkl-heat capacity
differ from the ordinary ones. Moreover, these differences alter according
to parity.

In this paper, we intend to investigate ideal Bose gas and blackbody
radiation with the Dunkl formalism. Both phenomena are of great importance
in the observation of quantum effects in nature, and so it is clear that it
should work in the Dunkl formalism as well. To this end we prepare the
manuscript as follows: In sect. 2, we introduce the Dunkl formalism which
will be used throughout the manuscript. Sect. 3 is devoted to the study of
the ideal Bose gas in the Dunkl formalism, where we construct the partition
function in the grand canonical ensemble and discuss the modifications to
the critical temperature as compared to the non-deformed case. In sect. 4,
we investigate several thermal quantities of blackbody radiation within the
Dunkl formalism. Finally, we briefly summarize our findings.

\section{Dunkl Formalism}

In Dunkl-quantum mechanics, the momentum operators are substituted with the
Dunkl-momentum operators \cite{Bilel1, Bilel2},%
\begin{equation}
\frac{\hbar }{i}D_{j}=\frac{\hbar }{i}\left[ \frac{d}{dx_{j}}+\frac{\theta }{%
x_{j}}(1-R_{j})\right] ,\text{ \ \ }j=1,2,3.
\end{equation}%
where $\theta $ is the {Wigner} parameter, and {$\mathbf{R}_{i}$} is
the reflection operator, {$\mathbf{R}_{i}\mathbf{=(-1)}^{x_{i}\frac{d}{dx_{i}}%
}$}, that obeys 
\begin{equation}
R_{i}f\left( x_{j}\right) =\delta _{ij}f\left( -x_{j}\right) ,\quad \quad
R_{i}\frac{d}{dx_{j}}=-\delta _{ij}\frac{d}{dx_{j}}R_{i},\quad \quad
R_{i}R_{j}=R_{j}R_{i}.
\end{equation}%
These substitutions lead to a modification of the Heisenberg algebra and
commutation relations 
\begin{equation}
\left[ D_{i},D_{j}\right] =0,\quad \quad \left[ x_{i},D_{j}\right] =\delta
_{ij}(1+2\mathbf{\theta }R_{\delta _{ij}}).
\end{equation}%
Upon introducing the following operators \cite{Marcelo} 
\begin{equation}
\overline{\phi }_{i}\leftrightarrow x_{i}=a_{i}^{+},\quad \text{and}\quad
\phi _{i}\leftrightarrow D_{x_{i}}=a_{i}+\frac{\theta }{a_{i}^{+}}\left(
1-R_{i}\right) ,  \label{eq}
\end{equation}%
where $\hat{a}_{i}^{+}$ and $\hat{a}_{i}$ are the ordinary creation and
annihilation operators, we {find that} they satisfy the following
commutation relations 
\begin{equation}
\left[ \phi _{i},\bar{\phi _{j}}\right] =\delta _{ij}(1+2\theta R_{i}).
\end{equation}%
One can therefore express the action of the modified operators on the
occupation states $\left\vert n\right\rangle \equiv \left\vert
n_{1},n_{2},n_{3},..\right\rangle $ as follows:%
\begin{eqnarray}
\bar{\phi}_{i}\left\vert n_{i}\right\rangle &=&\sqrt{n_{i}+1}\left\vert
n_{i}+1\right\rangle ,  \label{c} \\
\phi _{i}\left\vert n_{i}\right\rangle &=&\left[ \sqrt{n_{i}}+\frac{\theta }{%
\sqrt{n_{i}}}\left( 1-(-1)^{n_{i}}\right) \right] \left\vert
n_{i}\right\rangle .  \label{a}
\end{eqnarray}%
Using Eqs. (\ref{eq}) and (\ref{c}), (\ref{a}), the occupation number
operator $\mathcal{\hat{N}}_{i}$ can be written 
\begin{equation}
\bar{\phi}_{i}\phi _{i}=\mathcal{\hat{N}}_{i},
\end{equation}%
with 
\begin{equation}
\mathcal{\hat{N}}_{i}\left\vert n_{i}\right\rangle =\left[ n_{i}+\theta
\left( 1-(-1)^{n_{i}}\right) \right] \left\vert n_{i}\right\rangle .
\label{n}
\end{equation}

It is worth noting that, the eigenvalue of the modifid number operator
admits two cases ($n_{i}$: even or odd). For the even case, we get%
\begin{equation}
\mathcal{\hat{N}}_{i}\left\vert n_{i}\right\rangle =n_{i}\left\vert
n_{i}\right\rangle ,
\end{equation}%
while for the odd case, we have%
\begin{equation}
\mathcal{\hat{N}}_{i}\left\vert n_{i}\right\rangle =\left( n_{i}+2\theta
\right) \left\vert n_{i}\right\rangle .
\end{equation}

It's worthwhile to mention that the eigenvalues of the occupation number
operator $\mathcal{\hat{N}}_{i}$ separate into two sub-spaces according to
their parity. The even parity sub-space is the standard case. The notable
result is that only the odd eigenvalues are affected by the Wigner parameter.

\section{Ideal Bose gas in the Dunkl formalism}

Let us now consider the ideal Bose gas in the grand canonical ensemble. The
partition function of the system in the Dunkl formalism is given by%
\begin{equation}
Z^{D}=\sum_{\left\{ \mathcal{N}_{i}\right\} }e^{-\beta \sum_{\mathbf{i=0}}%
\mathcal{N}_{i}\left( \varepsilon _{i}-\mu \right) },  \label{z}
\end{equation}%
where $\mu $ is the chemical potential of the gas supposed at a temperature $%
T$. Here, $1/\beta =k_{B}T$, where $k_{B}$ denotes Boltzmann constant, and $%
\varepsilon _{i}$ represents the single-particle eigenenergy of the state "$%
i $". Using the definitions given in Eqs. (\ref{n}) and (\ref{z}), the partition function of the ideal Bose gas in the Dunkl formalism reads 
\begin{equation}
Z^{D}={\prod\limits_{i=0}^{\mathbf{\infty }}}\sum_{n_{i}=0}^{\mathbf{\infty }}\left\{ e^{-\beta
\left( \varepsilon _{i}-\mu \right) }\right\} ^{^{\left[ n_{i}+\theta \left(
1-(-1)^{n_{i}}\right) \right] }}.  \label{s}
\end{equation}%
After some algebra, equation (\ref{s}) simplifies to the following
expression 
\begin{equation}
Z^{D}={\prod\limits_{i=0}^{\mathbf{\infty }}}\frac{1+e^{-\beta \left(
1+2\theta \right) \left( \varepsilon _{i}-\mu \right) }}{1-e^{-2\beta \left(
\varepsilon _{i}-\mu \right) }}  \label{pa}
\end{equation}%
where the limit $\theta \longrightarrow 0$ is evidently the well-known partition function of the ideal Bose gas%
\begin{equation}
Z={\prod\limits_{i=0}^{\mathbf{\infty }}}\frac{1}{1-ze^{-\beta
\varepsilon _{i}}},
\end{equation}%
where $z=e^{\beta \mu }$ is the fugacity.

With the partition function (\ref{pa}) in hand, let us investigate the total
particle number. To this end, we use the thermodynamic definition: 
\begin{equation}
N=\frac{1}{\beta }\frac{\partial }{\partial \mu }\log Z^{D}{\bigg{|}_{T,V}}.
\label{N}
\end{equation}%
By substituting Eq. (\ref{pa}) into Eq. (\ref{N}), we may write the total
number of particles as%
\begin{equation}
N=N_{0}^{D}+N_{e}^{D},
\end{equation}%
where%
\begin{equation}
N_{0}^{D}=\frac{2}{z^{-2}-1}+\frac{(1+2\theta )}{z^{-(1+2\theta )}+1},
\end{equation}%
denotes the occupation number of the ground state, and
\begin{equation}
N_{e}^{D}={\sum_{i=1}^{\infty }}\left( \frac{2}{e^{2\beta \varepsilon
_{i}}z^{-2}-1}+\frac{(1+2\theta )}{e^{\beta (1+2\theta )\varepsilon
_{i}}z^{-(1+2\theta )}+1}\right) ,  \label{Ne}
\end{equation}%
specifies the occupation number of the excited states. In the thermodynamic
limit, { where in a  large volume ($V\rightarrow \infty$) the number of particles in the system is assumed to be large ($N\rightarrow \infty$) in such a way that the density is a constant  $\left(\frac{N}{V}=\text{const.}\right)$ 
\cite{walter,Lavagno,Elliott},} we use the conversion relation 
\begin{equation}
\sum_{\mathbf{i=1}}^{\infty }\rightarrow \frac{2\pi V}{h}\left( 2m\right)
^{3/2}\int_{0}^{+\infty }\sqrt{\varepsilon }d\varepsilon ,
\end{equation}%
to transform Eq. (\ref{Ne}) into%
\begin{equation}
\frac{N_{e}^{D}}{V}=\frac{2\pi }{h}\left( 2m\right) ^{3/2}\int \sqrt{%
\varepsilon }d\varepsilon \left( \frac{2}{e^{2\beta \varepsilon }z^{-2}-1}+%
\frac{(1+2\theta )}{e^{\beta (1+2\theta )\varepsilon }z^{-(1+2\theta )}+1}%
\right) ,  \label{int}
\end{equation}%
{where $h$ is the Planck constant, $m$ is the particle mass, and $V$ is the volume  of the system.} A straightforward calculation yields 
\begin{equation}
\frac{N_{e}^{D}}{V}=\frac{1}{\lambda ^{3}\sqrt{2}}\left[ g_{3/2}\left(
z^{2}\right) -\sqrt{\frac{2}{1+2\theta }}g_{3/2}\left( -z^{1+2\theta
}\right) \right] ,  \label{qq}
\end{equation}%
where $\lambda =\left( \frac{2\pi \hbar ^{2}\beta }{m}\right) ^{1/2}$ is the
de Broglie thermal wavelength, and%
\begin{equation}
g_{s}(z)=\frac{1}{\Gamma \left( s\right) }\int_{0}^{+\infty }\frac{x^{s-1}}{%
e^{x}z^{-1}-1}dx.
\end{equation}%
Thus, we can write the total number of particles as%
\begin{equation}
N=\frac{2}{z^{-2}-1}+\frac{(1+2\theta )}{z^{-(1+2\theta )}+1}+\frac{V}{%
\lambda ^{3}\sqrt{2}}\left[ g_{3/2}\left( z^{2}\right) -\sqrt{\frac{2}{%
1+2\theta }}g_{3/2}\left( -z^{1+2\theta }\right) \right] .  \label{13}
\end{equation}%
By utilizing the following property 
\begin{equation}
g_{s}(z)+g_{s}(-z)=2^{1-s}g_{s}(z^{2}),  \label{ply}
\end{equation}%
we rewrite Eq. (\ref{13}) as 
{
\begin{equation}
\frac{N}{V}=\frac{N_{0}^{D}}{V}+\frac{g_{3/2}\left( z,\theta \right)}{\lambda ^{3}} ,
\label{17}
\end{equation}}%
where%
\begin{equation}
g_{3/2}\left( z,\theta \right) =g_{3/2}(z)+g_{3/2}(-z)-\frac{1}{\sqrt{%
1+2\theta }}g_{3/2}\left( -z^{1+2\theta }\right) .
\end{equation}%
\begin{figure}[tbh]
\centering
\includegraphics[scale=1]{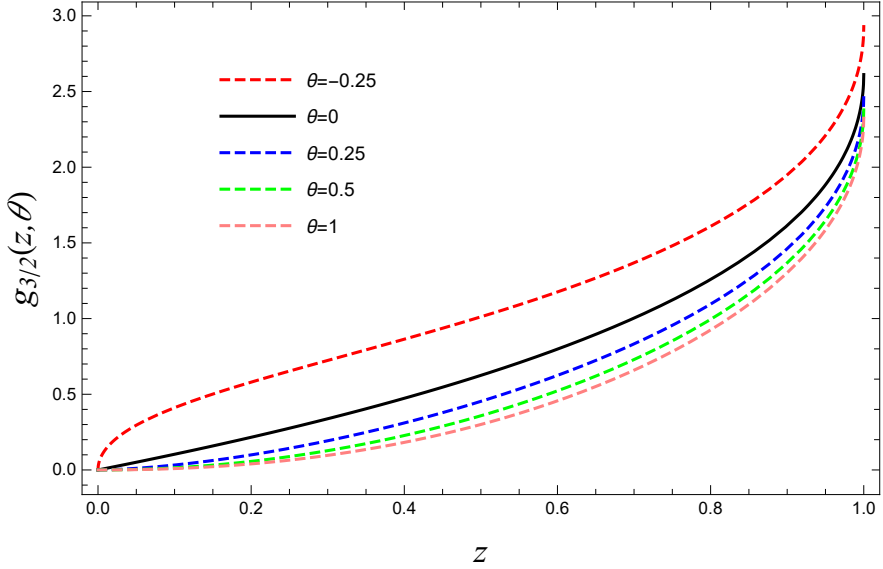}
\caption{The variation of $g_{3/2}\left( z,\protect\theta \right) $ via $z$
for different values of $\protect\theta .$}
\label{fig1}
\end{figure}

We notice on ({\ref{qq}}) that $N_{0}^{D}$ would be complex for $1+2\theta
<0 $. This non-physical behavior imposes a strong condition on the Wigner
parameter $\theta >-1/2$ which in turn means that the Dunkl formalism can
only describe ideal Bose gases in these regimes{.} Considering this
condition, we depict $g_{3/2}\left( z,\theta \right) $ versus $z$ via
different Wigner parameters in Fig. \ref{fig1}. We observe that the function
takes greater values when the Wigner parameter takes negative values, and
smaller values when it takes positive values.

\begin{figure}[tbph]
\centering\includegraphics[scale=1]{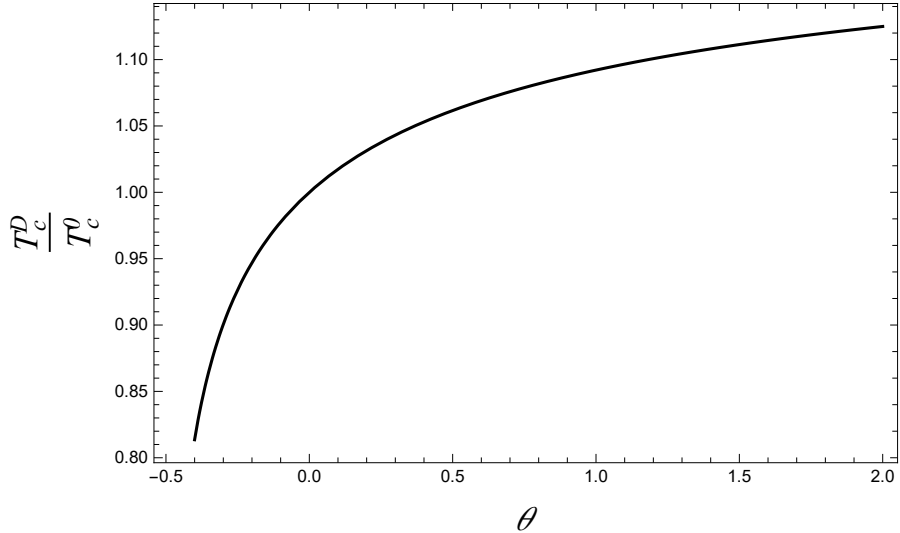}
\caption{Critical temperature ratio versus the Wigner parameter.}
\label{fig2}
\end{figure}

\bigskip We also note that, in the limit of $\theta \rightarrow 0$, Eq. ({%
\ref{17}}) reduces to the standard result:%
{
\begin{equation}
\frac{N}{V}=\frac{N_{0}}{V}+\frac{g_{3/2}\left( z\right)}{\lambda ^{3}} .
\end{equation}}%
To our best knowledge, two methods exist to compute the condensation (or
critical) temperature. The {first} method is to take $N_{0}^{D}=0$ in
Eq. ({\ref{17}}), and then, to work out $T_{c}^{D}$ \cite%
{Grossmann,Holthaus,Zeng,Cheng}. The second method is to calculate the
temperature at which the {specific} heat of the system reaches its
maximum value \cite{Kirsten,Toms}. In this manuscript, we use the fist
method. We get 
{
\begin{equation}
\frac{N}{V}=\frac{g_{3/2}(1)}{\lambda _{c}^{3}}\left[ 1+\frac{g_{3/2}(-1)}{g_{3/2}(1)}%
\left( 1-\frac{1}{\sqrt{1+2\theta }}\right) \right] ,  \label{nm}
\end{equation}}%
where $N$ is the actual total particle number. Hence, the Dunkl-condensation
temperature writes%
\begin{equation}
\frac{T_{c}^{D}}{T_{c}^{0}}=\left[ 1+\frac{g_{3/2}(-1)}{g_{3/2}(1)}\left( 1-%
\frac{1}{\sqrt{1+2\theta }}\right) \right] ^{-2/3},  \label{cr}
\end{equation}%
which clearly shows that the critical temperature is drastically modified by
the Wigner parameter. Indeed, only in the limit $\theta \rightarrow 0$, do
we obtain the standard condensation temperature $T_{c}^{0}=\frac{2\pi \hbar
^{2}}{mk_{B}}\left( \frac{N}{Vg_{3/2}\left( 1\right) }\right) ^{2/3}.$ 
This effect is better visualized by plotting the temperature ratio versus
the Wigner parameter (Fig. \ref{fig2}). We observe that the transition
temperature is smaller than $T_{c}^{0}$ when $-1/2<\theta <0$. On the other
hand, when $\theta $ takes on positive values, $T_{c}^{D}$ becomes greater
than $T_{c}^{0}$.

\begin{figure}[tbph]
\centering\includegraphics[scale=1]{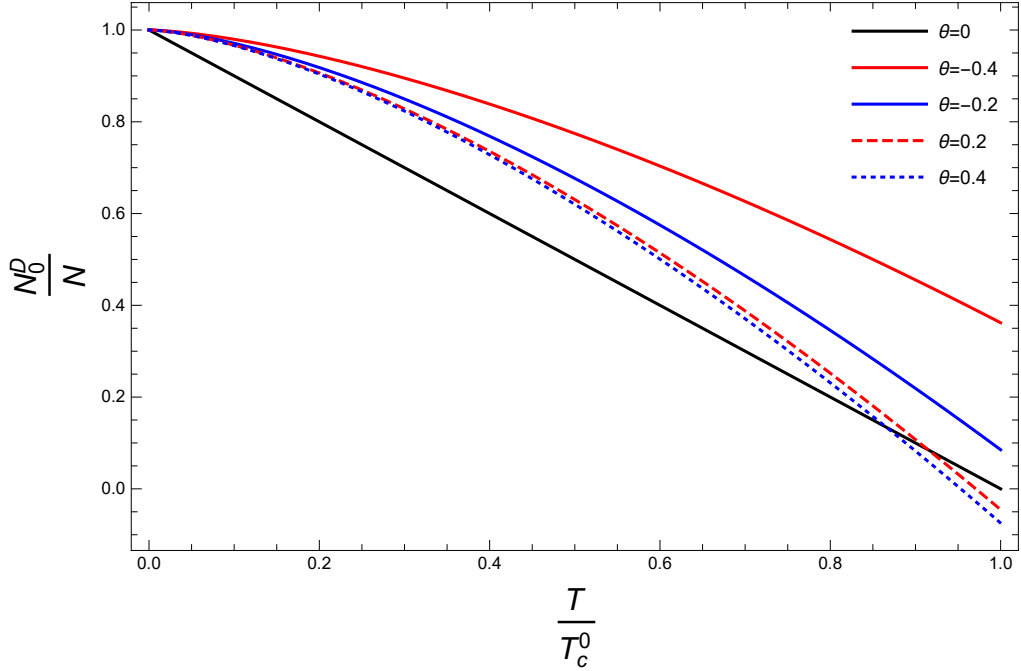}
\caption{Ground state population vs. the temperature ratio $T/T_{c}^{0}$ for
varying $\protect\theta $.}
\label{fig3}
\end{figure}

Finally, below the transition temperature, one may use Eqs. (\ref{17}) and (%
\ref{nm}) to derive the ground state population $\frac{N_{0}^{D}}{N}$ 
\begin{equation}
\frac{N_{0}^{D}}{N}=1-\left( \frac{T}{T_{c}^{D}}\right) ^{3/2},
\end{equation}%
In Fig. \ref{fig3}, we plot the ground state population $\frac{N_{0}^{D}}{N}$
as a function of $T/T_{c}^{0}$ for various $\theta $. 

\section{Blackbody radiation}

\begin{figure}[tbh]
\centering
\includegraphics[scale=1]{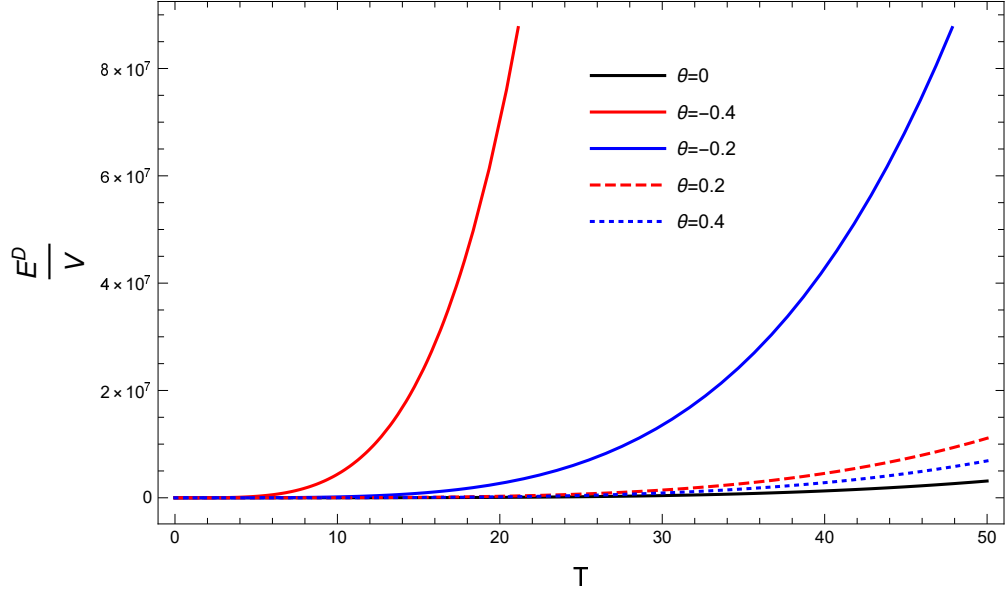}
\caption{Dunkl-corrected energy radiation per unit volume versus the
temperature for different values of $\protect\theta $.}
\label{fig4}
\end{figure}

In this section, we examine the blackbody radiation phenomena within the
Dunkl formalism. We start by expressing the mean occupation number of energy
state of photons, $N_{\varepsilon }$, in the Dunkl formalism for the
particular case of $\mu =0$. 
\begin{equation}
N_{\varepsilon }=\frac{2}{e^{2\beta \varepsilon }z^{-2}-1}+\frac{(1+2\theta )%
}{e^{\beta (1+2\theta )\varepsilon }z^{-(1+2\theta )}+1}.
\end{equation}%
By assuming $\varepsilon =\hbar \omega $, where $\omega $ is the frequency
of the photon, we write the number of photons within the frequency range of $%
\omega $ and $\omega +d\omega $ via: 
\begin{equation}
dN_{\varepsilon }=\frac{V}{\pi ^{2}c^{3}}\frac{\omega ^{2}d\omega }{%
e^{2\beta \hbar \omega }-1}+\frac{V(1+2\theta )}{2\pi ^{2}c^{3}}\frac{\omega
^{2}d\omega }{e^{\beta (1+2\theta )\hbar \omega }+1}.
\end{equation}%
Hence, the corresponding energy density to this radiation reads: 
\begin{equation}
\frac{dE^{D}}{V}=\frac{\hbar }{2\pi ^{2}c^{3}}\left[ \frac{2\omega ^{3}}{e^{%
\frac{2\hbar \omega }{\mathbf{k}_{B}T}}-1}+\frac{(1+2\theta )\omega ^{3}}{e^{%
\frac{(1+2\theta )\hbar \omega }{\mathbf{k}_{B}T}}+1}\right] d\omega .
\label{e}
\end{equation}%
After obtaining the generalized Planck radiation law in the framework of
Dunkl-statistics we integrate it over all frequencies. We find the deformed
total energy radiated by the cavity in the form of%
\begin{equation}
\frac{E^{D}}{V}=\frac{\sigma }{2c}T^{4}\left[ 1+\frac{7}{\left( 1+2\theta
\right) ^{3}}\right] ,  \label{en}
\end{equation}%
where $\sigma $ is the Stefan-Boltzmann constant. We observe that the
Dunkl-correction to the standard energy radiation arises with the second
term of Eq. (\ref{en}). We note that for $\theta =0$ the Dunkl-corrected
energy reduces to its conventional form, $\frac{E}{V}=\frac{4\sigma }{c}%
T^{4} $. In order to observe the effect of the Dunkl formalism, we depict
the radiated Dunkl energy by the cavity, $\frac{E^{D}}{V}$, versus the
temperature according to different deformation parameters in Fig. \ref{fig4}%
. 
We see that the radiated energy increases with the increasing temperature as
in the ordinary case. We also note that for a fixed value of temperature,
the energy radiation decreases when the deformation parameter grows.

\begin{figure}[tbh]
\centering
\includegraphics[scale=1]{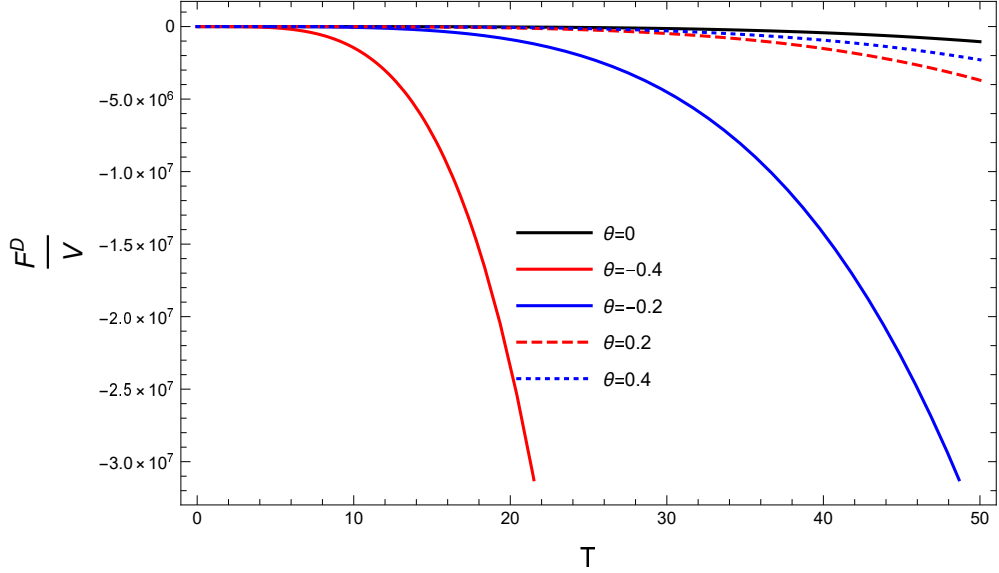}
\caption{Dunkl Helmholtz free energy per unit volume versus temperature for
different values of $\protect\theta $.}
\label{fig5}
\end{figure}

Next, we investigate several thermal quantities of the blackbody. First, we
study Dunkl-corrected Helmholtz free energy by employing 
\begin{equation}
F^{D}=-T\int \frac{E^{D}}{T^{2}}dT.  \label{fr}
\end{equation}%
After we substitute Eq. (\ref{en}) into Eq. (\ref{fr}), we arrive at,%
\begin{equation}
\frac{F^{D}}{V}=-\frac{\sigma }{6c}T^{4}\left[ 1+\frac{7}{(1+2\theta )^{3}}%
\right] .  \label{hel}
\end{equation}%
In the limit of $\theta \rightarrow 0$, we get the ordinary case result, $F=-%
\frac{4V}{3c}\sigma T^{4}$. In Fig. \ref{fig5}, we demonstrate the
Dunkl-Helmholtz free energy versus temperature. 

We observe that in all cases the Dunkl-Helmholtz free energy decreases
monotonically versus the increasing temperature. We also see that, for a
fixed value of $T$, the free energy function increases when the deformation
parameter $\theta $ grows.

The Dunkl-corrected entropy $S=-\left( \frac{\partial F}{\partial T}\right) $
is also of great interest. We indeed find the expression 
\begin{equation}
\frac{S^{D}}{V}=\frac{2}{3c}\sigma T^{3}\left[ 1+\frac{7}{(1+2\theta )^{3}}%
\right] ,
\end{equation}%
which shows, upon gathering (\ref{hel}) and (\ref{en}) that the
thermodynamic functions satisfy the standard relation 
\begin{equation}
F^{D}=E^{D}-TS^{D}.
\end{equation}%
The behavior of the Dunkl-corrected entropy versus temperature is depicted
in Fig. 
\begin{figure}[tbph]
\centering
\includegraphics[scale=1]{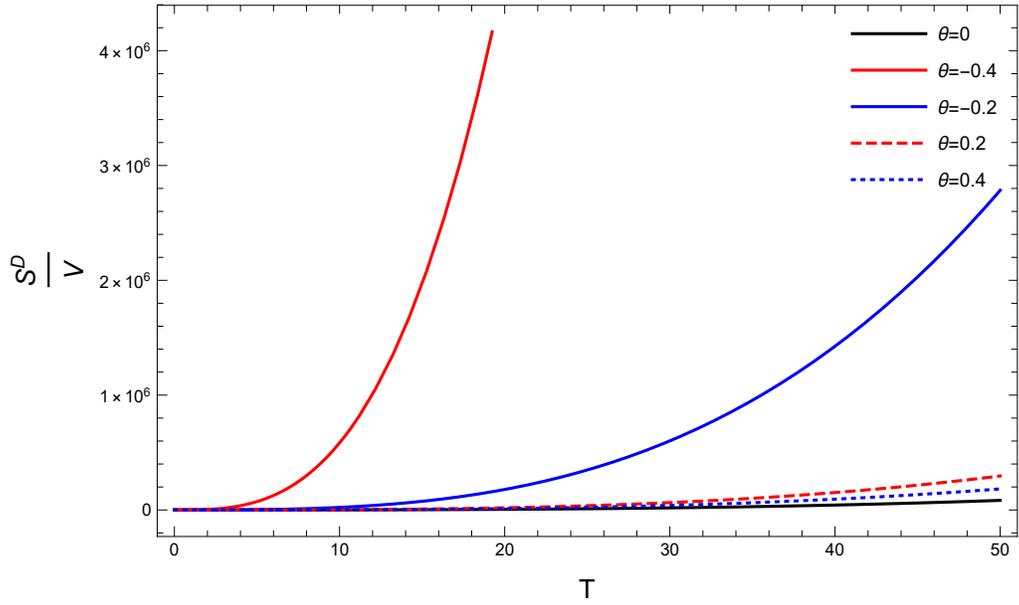}
\caption{The Dunkl entropy per unit volume as a function of temperature for
different values of $\protect\theta $.}
\label{fig6}
\end{figure}
We see that at low temperatures the effect of Dunkl formalism is not
observable. At high temperatures, this effect not only becomes significant
but there is also a saturation point for $\theta $ growing to infinity,
where one recovers the standard $T^{3}$ law with a different coefficient.

\begin{figure}[tbph]
\centering
\includegraphics[scale=1]{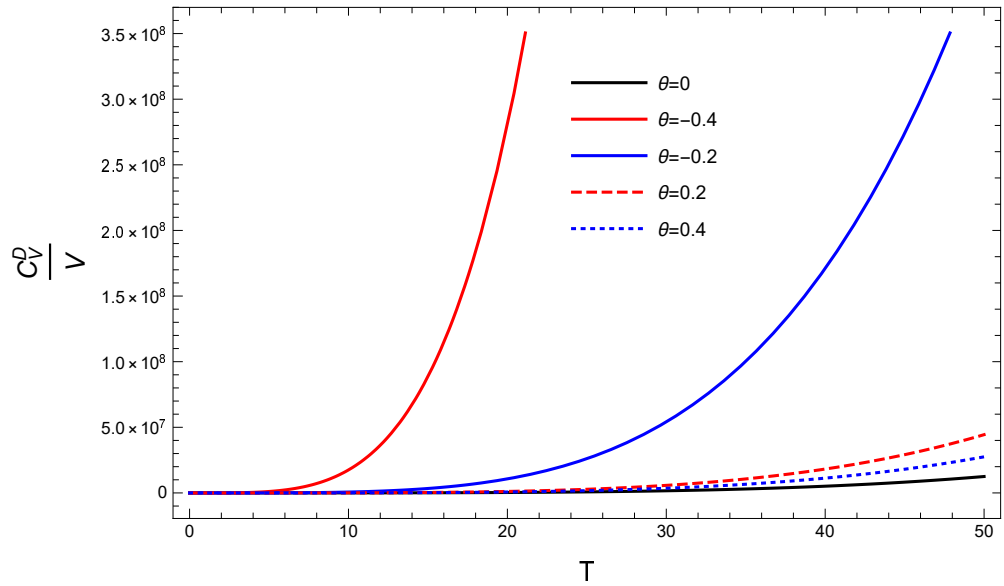}
\caption{Dunkl specific heat per unit volume as a function of temperature
for different values of $\protect\theta $.}
\label{fig7}
\end{figure}

Next, we study the Dunkl-corrected specific heat function at constant
volume, given by%
\begin{equation}
C_{V}^{D}=-\left( \frac{\partial E^{D}}{\partial T}\right) _{V}=\frac{2V}{c}%
\sigma T^{4}\left[ 1+\frac{7}{(1+2\theta )^{3}}\right] .  \label{DCV}
\end{equation}%
In the limit of $\theta \rightarrow 0$, Eq. (\ref{DCV}) gives the standard
result. We depict the behavior of the Dunkl-corrected specific heat versus
temperature in Fig. \ref{fig7}. 

We notice that the deformed specific heat function increases with increasing
temperature. This increase is greater for negative Wigner parameter.

Finally, we calculate the Dunkl-corrected pressure function of the model:%
\begin{equation}
P=-\left( \frac{\partial F}{\partial V}\right) _{T}.  \label{pre}
\end{equation}%
After substituting (\ref{hel}) into (\ref{pre}), we obtain 
\begin{equation}
P^{D}=\frac{\sigma }{6c}T^{4}\left[ 1+\frac{7}{\left( 1+2\theta \right) ^{3}}%
\right] .
\end{equation}
This yields the equation of state (EOS) in the form 
\begin{equation}
P^{D}V=\frac{E^{D}}{3},
\end{equation}
which shows that the EOS is invariant within Dunkl formalism.

\section{Conclusion}

In this manuscript, we employed Dunkl formalism to investigate two important
phenomena in physics. The ideal Bose gas in the grand canonical ensemble,
where we construct the partition function and derive the total number of
particles and the condensation temperature. A rigorous constraint appears on
the Wigner parameter, namely $\theta>-1/2$. Moreover, we found that the
deformed critical temperature becomes smaller or greater than the
non-deformed one according to the negative and positive values of the
deformation parameter.

In a second illustration, we examined the blackbody radiation within the
Dunkl formalism. We derived the total energy radiated and studied various
thermodynamic functions such as the Helmholtz free energy, the entropy, the
specific heat, and the pressure of the system. We observed first that the
Dunkl formalism leads to the same temperature dependence laws as the
non-deformed case but with coefficients depending on the deformation
parameter. These coefficients show a characteristic dependence on $\theta $.
Finally, we noticed that the equation of state is invariant in the Dunkl
formalism.

{Finally, it's worthwhile to mention that the occurrence of
Bose-Einstein condensation in an ideal gas is an elementary illustration of
a phase transition. A rigorous development of this phenomenon was given in
Refs \cite{Lewis,van,Robinson,Rocca,ZAGREBNOV}. M. van den Berg et al. studied the Bose-Einstein condensation in the perfect boson gas due to
the standard saturation mechanism in a prism of volume $V$ with
sides $V^{\alpha }$, $V^{\gamma }$, $V^{\delta }$
 with $\alpha \geq \gamma \geq \delta >0$\ and $\alpha +\gamma+\delta =1$ in \cite{Lewis} and showed that:

\begin{description}
\item[$I$]: if $\alpha <1/2$, above the critical density $\rho _{c}$, the condensation is formed only in the ground states.

\item[$II$]: if $\alpha =1/2$\textbf, the condensation is formed in an infinite number of single-particle states in a band with $\rho >\rho _{c}$.

\item[$III$]: if $\alpha >1/2$, non-extensive occupation is formed where $\rho _{0}=\rho
-\rho _{c}.$

\end{description}

Within the Dunkl formalism, the boson gas in rectangular parallelepiped of volume $V$ with sides of length $V^{\alpha }$, $V^{\gamma }$, $V^{\delta }$ may follow from a finite critical density the same way as in the conventional ideal Bose gas case. This case requires a thorough discussion which is currently under consideration and will be the subject of another study.}

\section*{Acknowledgments}

This work is supported by the Ministry of Higher Education and Scientific
Research, Algeria under the code: PRFU:B00L02UN020120220002. BCL is
supported by the Internal Project, [2023/2211], of Excellent Research of the
Faculty of Science of Hradec Kr\'alov\'e University.

\section*{Data Availability Statements}

The authors declare that the data supporting the findings of this study are
available within the article.

\end{document}